\documentclass[%
 aip,
 amsmath,amssymb,
 reprint,%
floatfix]{revtex4-1}

\usepackage{graphicx}
\usepackage{dcolumn}
\usepackage{bm}

\usepackage[utf8]{inputenc}
\usepackage[T1]{fontenc}
\usepackage{mathptmx}
\usepackage{etoolbox}

\makeatletter
\def\@email#1#2{%
 \endgroup
 \patchcmd{\titleblock@produce}
  {\frontmatter@RRAPformat}
  {\frontmatter@RRAPformat{\produce@RRAP{*#1\href{mailto:#2}{#2}}}\frontmatter@RRAPformat}
  {}{}
}%
\makeatother
\begin{document}

\preprint{AIP/123-QED}

\title[Slow interband recombination promotes an anomalous thermoelectric response of the $p-n$ junctions]{Slow interband recombination promotes an anomalous thermoelectric response of the $p-n$ junctions}
\author{Aleksandr S. Petrov}
 \email{petrov.as@mipt.ru}
\author{Dmitry Svintsov}%
\affiliation{ 
Laboratory of 2D Materials for Optoelectronics, Moscow Institute of Physics and Technology, Dolgoprudny 141700, Russia
}%

\date{\today}

\begin{abstract}
Thermoelectric effects in $p-n$ junctions are widely used for energy generation with thermal gradients, creation of compact Peltier refrigerators and, most recently, for sensitive detection of infrared and terahertz radiation. It is conventionally assumed that electrons and holes creating thermoelectric current are in equilibrium and share the common quasi-Fermi level. We show that lack of interband equilibrium results in an anomalous sign and magnitude of thermoelectric voltage developed across the $p-n$ junction. The anomalies appear provided the diffusion length of minority carriers exceeds the size of hot spot at the junction. Normal magnitude of thermoelectric voltage is partly restored if interband tunneling at the junction is allowed. The predicted effects can be relevant to the cryogenically cooled photodetectors based on bilayer graphene and mercury cadmium telluride quantum wells.
\end{abstract}

\maketitle

\begin{figure}
\includegraphics[width=\linewidth]{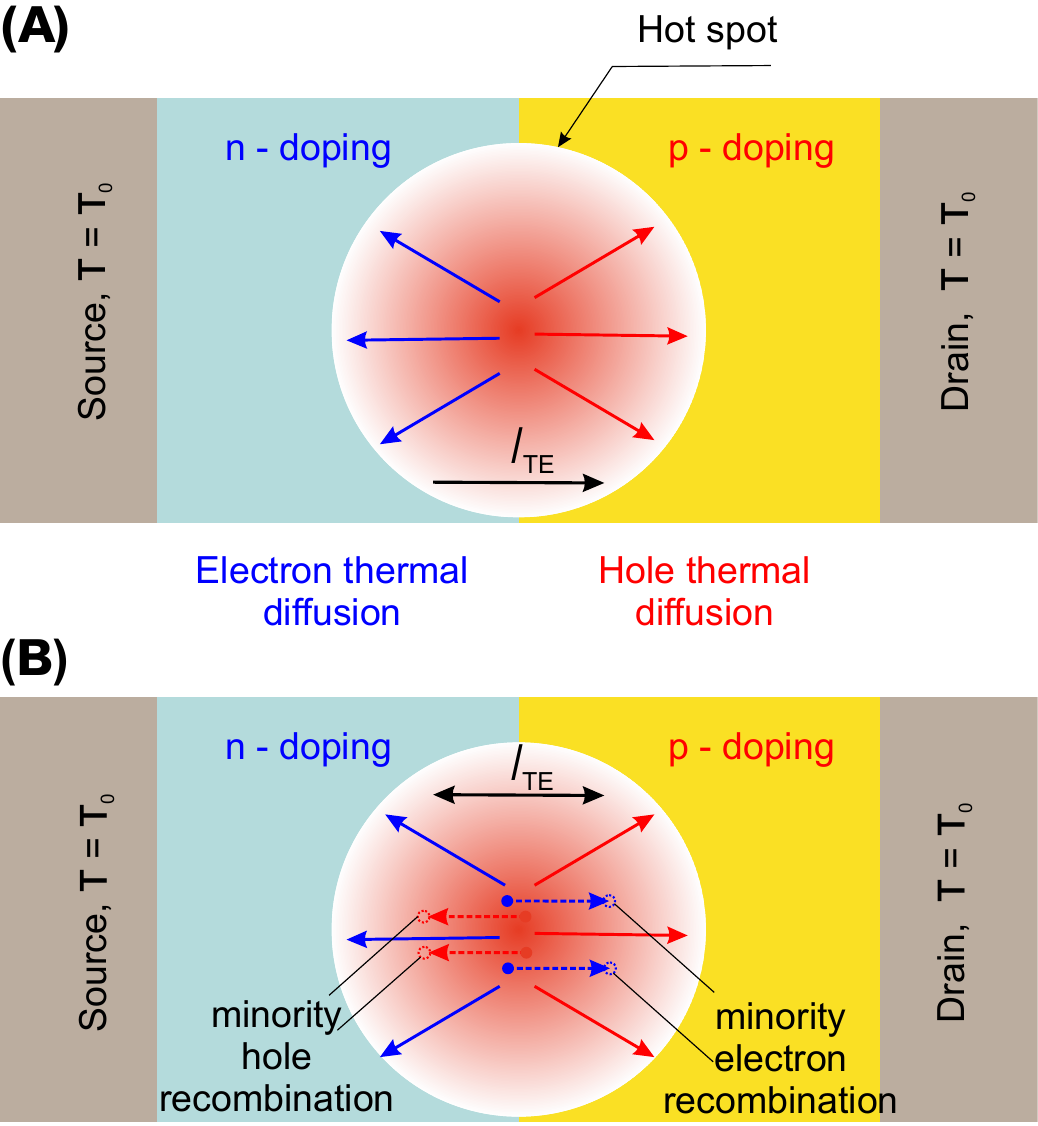}%
\caption{\label{fig-setups} 
Generation of thermoelectric current $I_{\rm TE}$ in a locally heated $p-n$ junction with cold leads (A) Conventional picture, where hot electrons diffuse to the source and hot holes diffuse to the drain, which results in rightwards thermoelectric current (B) Advanced picture, where minority carriers can penetrate into neighboring regions and exhibit thermal diffusion in opposite direction therein. Naturally, these minority carriers recombine with steady-state ones. The direction of thermoelectric current depends on the recombination rate
}%
\end{figure}

A heated $p-n$ junction with thermostatic contacts develops finite current under short circuit conditions and finite voltage under open circuit conditions. The sign of these current and voltage are typically predicted by assuming that both electrons and holes diffuse from the hot junction to the cold leads, as shown in Fig.~\ref{fig-setups} A. Considering the opposite charge of electrons and holes, this leads to the current $I_{\rm TE}$ directed rightwards and positive voltage at the drain. Such processes recently gained a new wave of interest in light of photodetectors based on two-dimensional materials with induced $p-n$ junctions~\cite{Rogalski_2d_PDs,Koppens_graphene_PDs}. The thermoelectric effects in these detectors are due to the local light-induced heating~\cite{Gabor2011}. They are especially relevant for terahertz and infrared radiation, where energy of the light quanta is insufficient to induce interband transitions~\cite{Castilla2019,Castilla2020,Titova2023a}. The effect of irradiation in such devices reduces thus to carrier heating via Drude absorption.

Several important questions can be raised to the electron-hole thermal diffusion picture in Fig.~\ref{fig-setups} A. Namely, what happens with electrons that attempt to diffuse from the hot junction rightwards to the cold drain? What happens to the holes that attempt to diffuse from the hot junction leftwards to the cold source? Once all the carriers have diffused from the hot junction to the cold contacts, how they can be replenished?

Attempting to answer these questions, we have to infer that the simple picture of Fig.~\ref{fig-setups} is valid under assumption of fast interband generation and recombination. Attempt of minority carrier thermal diffusion into majority region results in their interband recombination. Conversely, after the carriers leave the hot junction, the are replenished via interband generation. If we assume slow recombination of these minority carriers, the thermoelectric current can even change sign, as suggested in Fig.~\ref{fig-setups} B.

The success of thermoelectric picture with strong recombination in describing the photoresponse of graphene photodetectors stems from very intense Auger processes in this zero-gap semiconductor~\cite{Brida_Auger,Malic_Auger,Alymov2018a}. In gapped materials, such as bilayer graphene under transverse electric field~\cite{Gierz_Inversion} or mercury cadmium telluride quantum (HgCdTe) wells~\cite{Morozov_10THz_lasing}, the Auger processes are suppressed. The case of HgCdTe is particularly appealing as the gapped symmetric Dirac-type dispersion law increases the Auger recombination threshold~\cite{Alymov_HgCdTe}.

The interband non-equilibrium states formed upon forward electrical bias in $p-n$ junctions are well studied as they set the basis of laser diodes~\cite{Laser_diode_1,Laser_diode_2,pikusENG1965osnovy}. On the other hand, the fundamental literature on non-equilibrium states formed by thermal bias is scarce. Ref.~\onlinecite{Foster_Aleiner} has shown that thermal bias produces non-equilibrium electrons and holes, and suggested monolayer graphene as a suitable platform for observation of emerging corrections to thermoelectric coefficient. As the Auger scattering in monolayer graphene appeared strong, this study did not gain experimental confirmation~\cite{Kim_TE_Graphene}. Still, inclusion of electron-hole imbalance proved to be important for description of Coulomb drag in parallel graphene layers~\cite{Narozhny_hydrodynamics} or magnetotransport in charged fluids~\cite{alekseev2018nonmonotonic}. Ref.~\onlinecite{Polini_effective_Seebeck} introduced the effective Seebeck coefficient for graphene taking into account the interband imbalance caused by photoexcitation, but did not consider that temperature gradients cause imbalance by themselves. Apart from 2d materials, full set of drift-diffusion equations for thermoelectric transport in $p-n$ junctions with inclusion of non-equilibrium carriers was analyzed in~\cite{Gurevich_Pnj_Noneq_1,Gurevich_Pnj_Noneq_2}. Strong impact of recombination rate on Peltier cooling was predicted~\cite{Gurevich_Pnj_Noneq_1}, and efficient thermoelectric generators using non-equilibrium carriers were proposed~\cite{Span_Thermoelecrric_devices_eh_pairs}. Finally, inclusion of thermoelectric and recombination processes proved to be important to explain the asymmetric luminescence of Si micro-bridges~\cite{Bakan_Noneq_thermoelectric_Si}.

The present paper is devoted to the evaluation of thermoelectric voltage buildup in $p-n$ junctions with arbitrary rate of interband recombination. As the electron recombination rate varies from very high to very low values, the thermoelectric voltage changes sign. The contribution of minority carriers to the effective Seebeck coefficient becomes dominant for slow recombination and their low density. We finally show that finite tunneling transparency of the $p-n$ junction acts as a local source of recombination, thus partly restoring the conventional picture.


We consider the thermoelectric response of a symmetric $p$-$n$ junction (Fig. \ref{fig-setups}a). The temperature distribution along the junction $T(x)$ is symmetric with its maximum at the midpoint. Such local heating can be achieved by junction irradiation with focused laser beam~\cite{Gabor2011}. Alternatively, it is reached in split-gate junctions due to the near-field effects~\cite{Woessner_electrical_detection,Titova2023a}. We neglect the direct electron-hole generation which is possible only if the light quantum energy $h\nu$ is larger than the band gap $E_g$. Even in gapless materials, the interband electron-hole generation is suppressed at high doping with Fermi energies $\varepsilon_F > h \nu/2$ (Moss-Burstein effect)~\cite{Kono_spectroscopy}. We shall focus on determination of thermoelectric voltage $V_{\rm te}$ under open-circuit conditions. For small signals, the short-circuit current can be found by dividing $V_{\rm te}$ by resistance $R$.

Our model is based on drift-diffusion equations. The partial electron and hole currents are proportional to the gradients of quasi-Fermi levels $\partial_x\mathcal F_{e,h}$ and gradients of temperature
\begin{gather}
    \label{eq-material}
    j_{e,h} = \pm \sigma_{e,h}\partial_x\mathcal F_{e,h}/e \pm \alpha_{e,h}\partial_x T,
\end{gather}
where $\sigma_{e,h}(x)$ are the electron and hole dc conductivities, $\alpha_{e,h}(x)>0$ are their thermoelectric coefficients, and $e>0$ is the elementary charge. $\mathcal{F}(x)$ denotes the deviation of the local electrochemical potential from the equilibrium value $F_0(x)$; $\mathcal{F}_e(x)$ is counted upwards from $F_0(x)$ and describes non-equilibrium electron distribution in the conduction band; $\mathcal{F}_h(x)$ is counted downwards from $F_0(x)$ and describes non-equilibrium hole distribution in the valence band. The conductivity and the thermoelectric coefficient are calculated within the $\tau_p$-approximation to the Boltzmann kinetic equation, see Eqs.~(2) and (3) in Ref.~\cite{Titova2023a}.

We note here that the currents (\ref{eq-material}) include both drift and diffusion components, as they are proportional to the gradients of electrochemical potentials $\mathcal{F}_{e,h}(x)$. The potentials, in turn, include both chemical ($\varepsilon_F$) and electrical ($e\varphi$) components, $\mathcal{F}_{e}(x) = \varepsilon_{F,e}(x) - e\varphi(x)$ and  $\mathcal{F}_{h}(x) = \varepsilon_{F,h}(x) + e\varphi(x)$. Remarkably, the electric potential $\varphi(x)$ does not appear in equations separately from $\varepsilon_F$. As a result, the Poisson's equation relating the potential and charge density is not needed for modeling of thermoelectric response, least in the linear regime. This makes our results generic and independent on the dielectric environment, contact geometry, and even dimensionality of the thermoelectric material (be it 1d, 2d or 3d).

During their thermal diffusion, electrons and holes recombine with the rate $\mathcal{R}(x)$ according to the continuity equation
\begin{equation}
    \label{eq-continuity}
    \partial_x j_{e,h} = (\mp e)\mathcal{R}.
\end{equation}
Under open circuit conditions, the net current of electrons and holes is zero
\begin{equation}
    \label{eq-Jtot}
    j_e + j_h = j_{tot} \equiv 0.
\end{equation}
Finally, electrons and holes are equilibrated at the source and drain contacts, which we position symmetrically at $x=-L/2$ and $x=L/2$. A formal representation of this fact is equality of conduction and valence bands' quasi-Fermi levels at $x=\pm L/2$, and their equality to the generated thermoelectric voltage:
\begin{equation}
    \label{eq-bound-contact}
    \mathcal{F}_{e,h}(-L/2) = \pm \frac{eV_{ph}}{2}; \mathcal{F}_{e,h}(L/2) = \mp \frac{eV_{ph}}{2};
\end{equation}

For small deviations $\mathcal{F}_{e,h}$ the recombination rate can be written as
\begin{equation}
    \mathcal{R}(x) = R_0 (\mathcal{F}_e(x) + \mathcal{F}_h(x))/T_0,
\end{equation}
where $R_0 [1/\left(\mathrm{s}\cdot\mathrm{cm}^2\right)]$ is the recombination constant, and the electrochemical potential deviations are normalized by the base temperature $T_0$ (in energy units). For illustrative purposes we adopt a simple "recombination time" approximation when $R_0 = n_R/\tau_R$, where $n_R = m T_0/2\pi\hbar^2$ is characteristic carrier density and $\tau_R$ is the recombination time.

The formulated model is readily solvable in the conventional case of very fast recombination, $\mathcal{R} \to \infty$. Electrons and holes share the same quasi-Fermi level, which means $\mathcal{F}_e(x) \equiv -\mathcal{F}_h(x)$. This situation is illustrated by Fig.~\ref{fig-setups}b. The thermoelectric photovoltage $eV_{ph} = \mathcal{F}_e(L/2)-\mathcal{F}_e(-L/2)$ is easily expressed from Eqs.(\ref{eq-continuity})-(\ref{eq-Jtot}) as
\begin{equation}
\label{eq-te-inf-r}
    V_{ph}^{\infty} = \int\limits_{-L/2}^{L/2} \frac{\alpha_e(x)-\alpha_h(x)}{\sigma_e(x)+\sigma_h(x)}\partial_x T\,dx,
\end{equation}
where the superscript $\infty$ marks infinite recombination rate. The fraction under the integration sign is the negative of Seebeck coefficient $-S^{\infty}(x)$. 


The opposite limiting case of weak recombination, $\mathcal{R} \to 0$, almost not discussed in the literature, is also simply addressed with Eqs.~(\ref{eq-material}) -- (\ref{eq-Jtot}). We express the local voltage $[\mathcal{F}_e(x) + \mathcal{F}_h(x)]/2$ from (\ref{eq-material}) considering $\mathcal{R} \equiv 0$. Integration of this quantity between source and drain yields $eV_{\rm ph}$ by definition. From the other hand, it yields 
\begin{equation}
\label{eq-te-zero-r}
    V_{ph}^{0} = \int\limits_{-L/2}^{L/2} \left[\frac{\alpha_e(x)}{\sigma_e(x)} - \frac{\alpha_h(x)}{\sigma_h(x)}\right] \partial_x T\,dx,
\end{equation}
where the quantity in the brackets can be called the Seebeck coefficient for the absence of recombination $-S^{(0)}(x)$. Remarkably, $S^{(0)}$ usually has the opposite sign to $S^{\infty}$, which we checked numerically, so we may argue that $S^{(0)}$ is governed by the minority carriers, while $S^{\infty}$ is governed by the majority ones.

We perform further analysis for the stepwise model of the $p-n$ junction, where the length of the transition region is negligibly small compared to the hot spot size. In that case, the photovoltages for fast and slow recombination depend only on the local junction heating $\delta T(0) = T(0)-T_0$ via
\begin{gather}
\label{eq-sharp-te-inf-r}
    V_{ph}^{\infty} = -2S_n^\infty \delta T(0) = 2S_p^\infty \delta T(0),\\
\label{eq-sharp-te-zero-r}
    V_{ph}^{0} = -2S_n^0 \delta T(0) = 2S_p^0 \delta T(0),
\end{gather}
where $S_n = -S_p$ are the Seebeck coefficients of the $n$- and $p$-doped regions. 

Within the stepwise junction model, the linearized drift-diffusion equations in each region have constant coefficients before $\mathcal{F}_{e,h}$ and $\mathcal{F}''_{e,h}$ and can be solved analytically. As the simplest approximation to the transport at the junction $x=0$, we adopt the continuity of currents and quasi-Fermi levels $j_{e,h}(-0) = j_{e,h}(+0)$, $\mathcal{F}_{e,h}(-0) = \mathcal{F}_{e,h}(+0)$. Continuity of partial currents implies small recombination rate within the junction, which can always be achieved by reducing its length. Continuity of quasi-Fermi level implies small resistivity of the junction, as compared to the bulk regions. This assumption will be released in the following.

Finally, we assume parabolic model form of the temperature distribution
\begin{equation}
\label{eq-parabolic}
    \Delta T(x) = T(x) - T_0 = \left[\left(\frac{2x}{L}\right)^2-1\right]\delta T.
\end{equation}

\begin{figure}[h!]
\includegraphics[width=\linewidth]{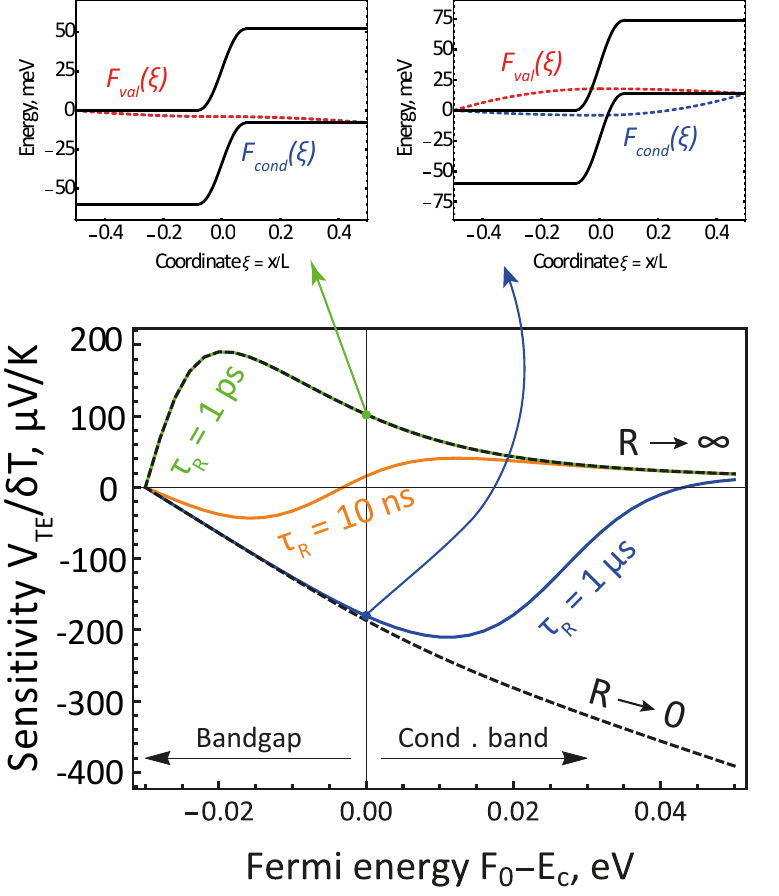}%
\caption{\label{fig-Gtun=0_tauR} (Top panel) Calculated band positions for the case $F_0 = 0\,$eV and different recombination times; the junction is depicted smooth for illustrative purposes (Bottom panel) Calculated photovoltage dependence on the doping level for different e-h recombination time. Black dashed curves correspond to limits of "fast recombination" (\ref{eq-te-inf-r}) and "slow recombination" (\ref{eq-te-zero-r}). Solid line  are the solutions for intermediate values of the recombination time $\tau_R$. The tunneling current at the junction is neglected. Parameters of calculation: $E_g = 0.06\,$eV, $m=0.03m_e$, $T_0 = 77\,$K, $\delta T = 77\,$K, device length $L=1\,\mu$m. Tunnel conductivity at the junction $\mathcal{G}_{tun}$ is absent.}
\end{figure}

The resulting expression for the photovoltage in the stepwise model of the $p-n$ junction and for parabolic temperature profile reads as follows:
\begin{equation}
    \label{eq-Vphoto-general-no-tun}
    V_{ph} = \left[1-g\left(\frac{\kappa L}{2}\right)\right] V_{ph}^{\infty} + g\left(\frac{\kappa L}{2}\right) V_{ph}^{0}.
\end{equation}
Here, $V_{ph}^{\infty}$ and $V_{ph}^0$ are given by Eqs.~\ref{eq-sharp-te-inf-r} and \ref{eq-sharp-te-zero-r}, respectively; the weight function
\begin{equation}
    g(\xi) = \frac{1}{\xi^2} - \frac{1}{\xi^2 \cosh\xi}
\end{equation}
quantifies the strength of recombination. The argument of the weight function is $\kappa L/2$, where $\kappa$ is the inverse diffusion length of non-equilibrium carriers
\begin{equation}
    \label{eq-kappa}
    \kappa = \sqrt{\frac{e^2n_R}{T_0 \tau_R}\rho_{tot}}.
\end{equation}
Here $\rho_{tot} = \left(\frac{1}{\sigma_e} + \frac{1}{\sigma_h}\right)$ is the total resistivity of the sample. Apparently, fast recombination corresponds to $\kappa L\gg 1$ and slow recombination -- to $\kappa L\ll 1$.


The photovoltage given by Eq.~\ref{eq-Vphoto-general-no-tun} is plotted in Fig.~\ref{fig-Gtun=0_tauR} as a function of equilibrium doping. The latter is parametrized by Fermi level reckoned from the conduction band edge $F_0 - E_c$, the values $F_0 - E_c <0$ correspond to the junction of non-degenerate semiconductor regions, while $F_0 - E_c>0$ correspond to degenerately doped sides. We adopt the parameter values typical for narrow-gap infrared materials like graphene bilayer~\cite{Falko_Electrons_GBL} and HgCdTe quantum wells~\cite{Zholudev_HgTe_spectroscopy}, $E_g = 60$ meV and $m^* = 0.03 m_0$. 

We observe that the sensitivity curves $V_{\rm te}/\delta T$ lie between the two limiting cases: Eq.~(\ref{eq-sharp-te-inf-r}) (black dashed curve) and Eq.~(\ref{eq-sharp-te-zero-r}) (red dashed curve). At high doping, the limit of fast recombination is favorably achieved, as the minority carrier entering the degenerately doped region has a large number of 'recombination partners'. Strong deviations from the conventional picture of fast recombination are favorably observed for weakly doped junctions, where minority carrier diffusion lengths are the largest. Remarkably, the thermoelectric voltage for slow recombination does not saturate for $F_0 - E_c \gg 1$, thus strongly violating the Mott's law in the metallic limit.

The narrow-gap semiconductors often display ambipolar transport ensured by large tunneling transparency of the $p-n$ junctions. The most recognized example is Klein tunneling in single-layer graphene~\cite{Klein_SLG}, though gapped 2d systems also display strong interband tunneling~\cite{Klein_BLG,Kvon_HgTe1,Kvon_HgTe2}. To account for these effect, we modify the boundary conditions for partial currents at $x=0$. Namely, we assume that incident carriers can either attempt to tunnel through band gap or traverse over the barrier, depending on their energy. Formally
\begin{gather}
    \label{eq-bound-junction}
    j_e(-0) = j_{tun} + j_e^{th}, \\
    j_h(+0) = j_{tun} + j_h^{th}.
\end{gather}
In line with previous theories of short tunnel transparent junctions~\cite{Crowell1966,Card1977}, we assume that the thermionic and tunneling currents are proportional to the discontinuities in the respective quasi-Fermi levels,
\begin{gather}
    \label{eq-jtun}
    j_{tun} = -\mathcal{G}_{tun}\left(\mathcal{F}_e^{(l)}(-0) + \mathcal{F}_h^{(r)}(+0)\right); \\
    \label{eq-jth}
    j^{th}_{e,h} = \pm\mathcal{G}_{th}\left(\mathcal{F}_{e,h}^{(r)}(+0) - \mathcal{F}_{e,h}^{(l)}(-0)\right),
\end{gather}
where $\mathcal{G}_{tun}$ and $\mathcal{G}_{th}$ are the tunnel and thermionic conductivities, respectively. They can be evaluated microscopically known the barrier height and energy dependence of interband transmission probability $\mathcal{T}(E)$. We shall refrain from these calculations and examine the qualitative changes in thermoelectric voltage $V_{\rm te}$ with variations of conductance. The only fact we take for known is the absence of tunneling for non-overlapping conduction and valence bands, i.e. $\mathcal{G}_{tun} \equiv 0$ for $F_0 < E_c$. 

\begin{figure}[h!]
\includegraphics[width=\linewidth]{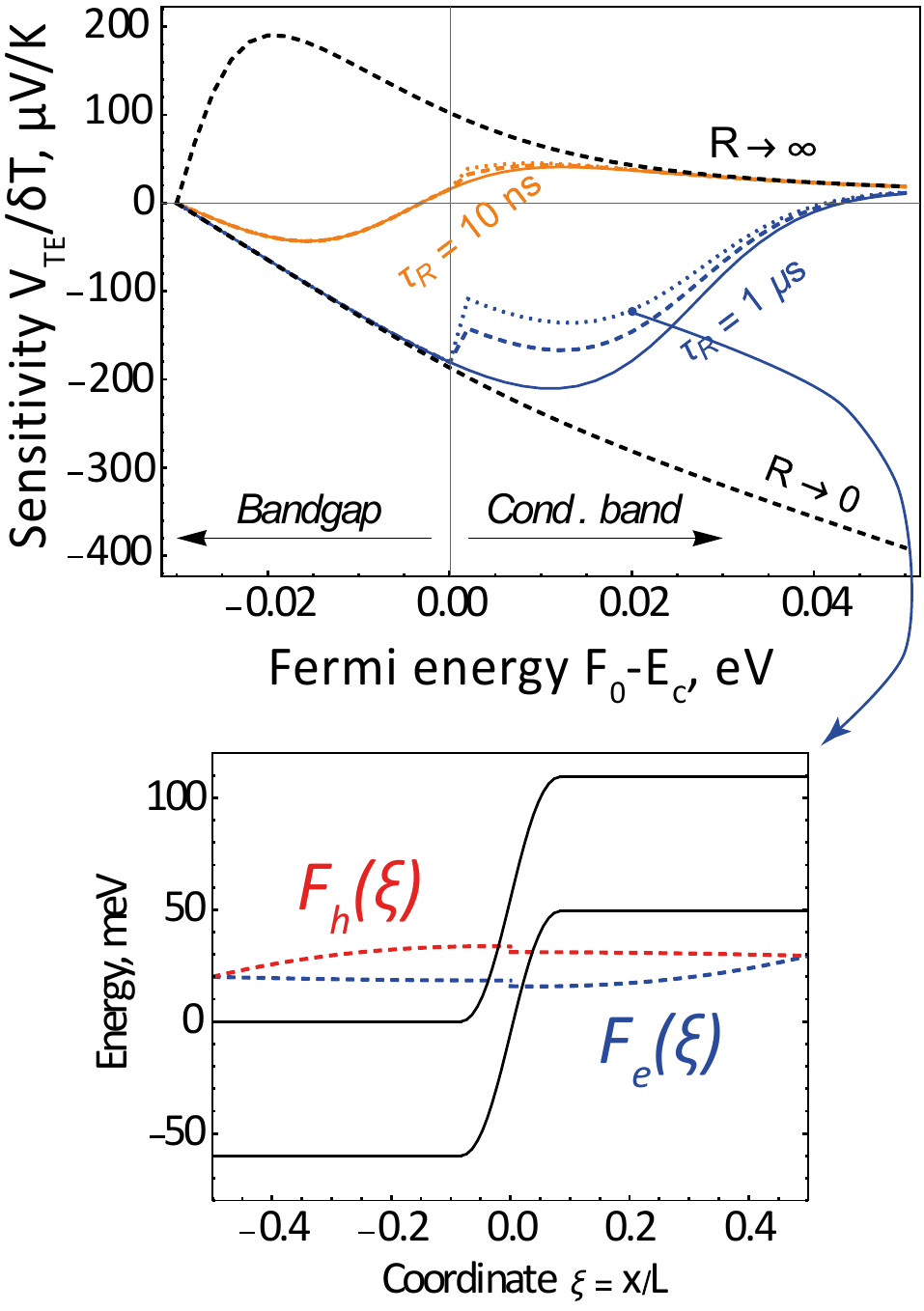}%
\caption{\label{fig-Gtun!=0_tauR} (Top panel) Calculated photovoltage dependence on the doping level for different e-h recombination time and tunneling conductivity. Solid lines correspond to no tunneling, colored dashed line -- $\mathcal G_{tun} = 0.2 \mathcal G_{th}$, colored dotted lines --  $\mathcal G_{tun} = 0.4 \mathcal G_{th}$. The other parameters are the same as on Fig.~\ref{fig-Gtun=0_tauR} (Bottom panel) Calculated band positions for the case $F_0 = 0.02\,$eV; the junction is depicted smooth for illustrative purposes}
\end{figure}

The doping dependencies of thermoelectric voltage in tunnel-transparent $p-n$ junction are shown in Fig.~\ref{fig-Gtun!=0_tauR} for different tunneling conductivities,varying between zero and $0.4\mathcal{G}_{\rm th}$. 
Expectantly, the picture for $F_0 < E_c$ remains the same as the valence and conduction bands do not overlap. For higher doping levels, the tunneling provides an additional recombination pathway for nonequilibrium carriers from either sides of the junction. This leads to the photovoltage curve shift towards the fast recombination limit.

We finally show that the recombination times required for observation of anomalous thermoelectric response are realistic. Let us assume that the hot spot is formed by a focused infrared laser with $L \approx 10$ $\mu$m. A non-equilibrium carrier can traverse across this spot without recombination provided $L < v_0 \sqrt{\tau_p \tau_R}$, where $v_0$ is the characteristic (Fermi or thermal) velocity and $\tau_p$ is the momentum relaxation time. Taking $v_0 \approx 10^{5}$ m/s and $\tau_p \approx 1$ ps, we arrive at an estimate $\tau_R > 10$ ns. Such recombination times are readily achieved at liquid nitrogen temperature in sub-critical ($d<6$ nm) HgTe quantum wells~\cite{Alymov_HgCdTe}. Even more favorable conditions for the recombination-free thermoelectric response appear in $p-n$ junctions created by electrical biasing of split-gates above the 2d materials~\cite{Woessner_electrical_detection,Titova2023a}. In such structures, the hot spot is localized within the inter-gate gap, which readily scales down to hundreds of nanometers.

To conclude, we have shown that suppression of the interband recombination in $p-n$ junctions leads to an anomalous flipping of the thermoelectric voltage. The effect takes place if the diffusion length of the minority carriers exceeds the size of the hot spot. Finite tunneling transparency of the junction acts as a localized recombination center and partly restores the normal thermoelectric response.

{\bf Acknowledgment.} The authors thank Dmitry Mylnikov for the discussions that stimulated the present work.

{\bf Funding.} The work was supported by grant \# 21-79-20225 of the Russian Science Foundation.

{\bf Conflict of interest.} The authors declare that they have no conflict of interest.

\nocite{*}
\bibliography{bibliography}

\end{document}